\documentclass[two column]{aastex631}

\usepackage{enumitem}
\usepackage{amsmath}
\usepackage{graphicx}
\usepackage{fancyhdr}
\usepackage{comment}
\usepackage[mathlines]{lineno}
\usepackage{threeparttable}
\usepackage{multirow}
\usepackage{CJKutf8}

\usepackage{hyperref}
\hypersetup{colorlinks=true, citecolor=blue, linkcolor=blue, urlcolor=blue}
\defcitealias{2022Nature_Xiang}{XR22}

\begin{document}
\title{The Epoch of the GSE Merger: Insights from the Splash and Thick Disk Age Distributions}

\newcommand{\KIAA}{\affiliation{Kavli Institute for Astronomy and Astrophysics, Peking University, Beijing 100871, China}}
\newcommand{\DoA}{\affiliation{Department of Astronomy, School of Physics, Peking University, Beijing 100871, China}}
\newcommand{\UCAS}{\affiliation{School of Astronomy and Space Science, University of Chinese Academy of Sciences, Beijing 100049, China}}
\newcommand{\NAOC}{\affiliation{National Astronomical Observatories, Chinese Academy of Sciences, Beijing 100101, China}}

\author{Sihan Lu}\DoA\KIAA
\author[0000-0002-7727-1699]{Yang Huang}\UCAS\NAOC
\author{Qikang Feng}\DoA\KIAA
\author[0000-0002-7727-1699,sname='Zhang']{Huawei Zhang}\DoA\KIAA

\correspondingauthor{Yang Huang and Huawei Zhang}
\email{huangyang@ucas.ac.cn and zhanghw@pku.edu.cn}

\begin{abstract}
The epoch of the Gaia-Sausage-Enceladus (GSE) merger, the last major merger experienced by the Milky Way, is crucial for reconstructing the Galaxy’s evolutionary history. This event dynamically heated the disk, scattering some stars into the halo and producing the so-called Splash. Yet the observed age distribution of the Splash is systematically older than that of the thick disk from which it is thought to originate, posing a puzzling inconsistency.
In this work, we show that this apparent discrepancy can be naturally explained once stellar age uncertainties are taken into account. The GSE truncated the intrinsic age distribution of the Splash at the merger epoch, while measurement errors introduce an Eddington-like bias that systematically shifts the truncated distribution toward older ages. Importantly, the magnitude of this shift depends on both the thick-disk star formation history and the merger epoch, thereby offering a new way to constrain the timing of the merger.
By combining the observed Splash age distribution with the peak of thick-disk star formation, we infer a GSE merger epoch of $ 10.1^{+0.2}_{-0.2}$ Gyr ago, providing one of the tightest constraints to date. This result offers new insight into the Milky Way’s accretion history and opens a path toward more robust reconstructions of its early evolution.
\end{abstract}


\section{Introduction} \label{sec:intro}
Unraveling the structure and evolutionary history of the Milky Way remains a central challenge in Galactic astronomy. With the advent of large-scale stellar surveys such as Gaia \citep{GAIA2016}, LAMOST \citep{2015RAALAMOST}, GALAH \citep{GALAH2018}, and APOGEE \citep{APOGEE2017}, unprecedented volumes of high-precision astrometric, photometric and spectroscopic data have become available. The kinematic and chemical information provided by these surveys enables the reconstruction of the Milky Way’s assembly history from its stars—long-lived “fossils” that preserve imprints of their formation conditions \citep[e.g.,][]{Imig_2023bimodal,2024Archaeology}.
In addition, stellar ages can be estimated using several complementary techniques: isochrone fitting for large samples \citep[e.g.,][]{Yi_2001isochrones, parsec_isochrones}, various machine-learning approaches \citep[e.g.,][]{Huang_RC_2020,astroNNage2022,Wang_2023_RGB}, and asteroseismology \citep[e.g.,][]{asteroseismology2020}. Together, these advances have opened a new window onto the formation and assembly history of the Milky Way. \par

A major focus has been on identifying the signal of past merger events that shaped the Milky Way’s present-day structure. Numerous stellar substructures have been identified, many of which are interpreted as remnants of past accretion events during the assembly of the Milky Way \citep[e.g.,][]{helmi_streams, Deason_2013_broken_stellar_density, Sequoia2019, Naidu_substructure2020, Malhan_wukong2021}. Among these events, the Gaia-Sausage-Enceladus (GSE) merger has had the most profound impact. The debris of this event was revealed with the release of Gaia \citep{Gaia_DR2, GSE_time2018, Helmi2018}, showing that the Milky Way’s disk was dynamically heated during the encounter \citep{2019_heat_disk,2019A&A_heat_disk}. In particular, a fraction of stars originally on near-circular disk orbits were scattered into the inner halo on lower-angular-momentum orbits, producing the so-called ‘Splash’ population \citep{splash2020}.\par 

Building on these discoveries, several studies have attempted to determine the epoch of the GSE merger. By comparing the observed velocity ellipsoid with zoom-in simulations, \citet{GSE_time2018} inferred that this major merger occurred between 8 and 11 Gyr ago. This estimate broadly agrees with constraints based on the metallicities and ages of halo stars, which suggest a merger time of approximately 10 Gyr ago \citep{Helmi2018,GSE_time_10Gyrs}. However, the relatively broad range of these estimates indicates that the timing of the GSE merger remains uncertain, limiting our ability to precisely reconstruct the Galaxy’s main accretion event.
The age distribution of the Splash population provides an additional potential constraint on the epoch of the GSE merger. Using stellar age dating, \citet{2022Nature_Xiang} found that splashed disk stars can be as old as $\sim$11 Gyr. In this context, the Splash population is generally thought to originate from the pre-merger thick disk. However, its age distribution appears systematically older than that of typical thick-disk stars \citep{2020MNRASselect_splash,2022Nature_Xiang}, posing a puzzling and seemingly counterintuitive result.\par 

The apparent paradox described above can in fact be explained by an observational bias akin to the Eddington-like bias \citep{1913_eddington_error}. Specifically, if the GSE merger truncated the intrinsic age distribution of the Splash while the thick disk continued forming stars, age uncertainties naturally shift the truncated distribution toward older ages. Crucially, the magnitude of this shift depends on both the merger epoch and the star formation history of the thick disk. Consequently, the observed difference between the age distributions of the Splash and thick-disk populations provides a new way to constrain the epoch of the GSE merger.\par

In this Letter, we use a large stellar catalog with full phase-space information, chemical abundances, and accurate stellar ages to investigate the timing of the GSE merger. 
The structure of this paper is as follows. Section~\ref{sec:2} describes the adopted sample and the selection of Splash and thick-disk populations. Section~\ref{sec:3} presents the constraint on the merger time based on the age distributions of the two populations. Section~\ref{sec:4} discusses the results and summarizes our main conclusions.

\section{Data} \label{sec:2}
\subsection{Age sample} \label{sec:2.1}
The data used in this study were drawn from \citet[][hereafter  \citetalias{2022Nature_Xiang}]{2022Nature_Xiang}\footnote{\url{https://keeper.mpdl.mpg.de/d/019ec71212934847bfed/}}, a catalog constructed by combining spectroscopic information from LAMOST DR7 \citep{2012RAALAMOST,2012RAALAMOST_survey} with astrometric information from Gaia eDR3 \citep{2021A&AGaiaeDR3}. In this study, subgiant stars were selected from the effective temperature–absolute magnitude diagram, yielding a parent sample of 247,104 stars. We adopt this catalog because it provides a large sample with high-precision stellar parameters, including 3D positions, 3D velocities, orbital dynamic parameters, detailed chemical abundance ratios, and reliable stellar age estimates, as well as the errors of these parameters. Stellar ages were derived through Bayesian isochrone fitting using the Yonsei--Yale stellar evolution models \citep{2004_YY_isochornes}, with 73\% of them having a relative error of less than 10\% and 99\% having a relative error of less than 20\%. The original sample has already undergone extensive quality control, including the removal of binary stars, sources with Gaia re-normalized unit weight error (RUWE) greater than 1.2, and stars with signal-to-noise ratios smaller than 20. We therefore apply only additional astrometric quality criteria, retaining 229,714 stars with well-measured radial velocities, proper motions and parallaxes:
$\texttt{vlos\_err} < 50 \text{ km s}^{-1}$;\quad
$\texttt{pmra\_error}/\texttt{pmra} < 0.2$;\quad $\texttt{pmdec\_error}/\texttt{pmdec} < 0.2$;\quad
$\texttt{parallax\_error}/\texttt{parallax} < 0.2$.\par

\begin{figure}[t]
    \centering
    \plotone{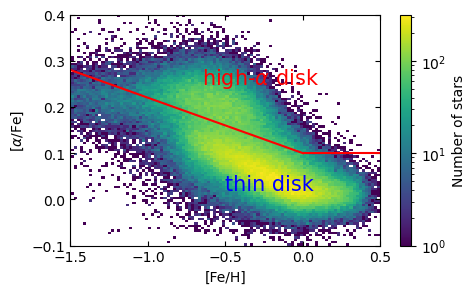}
    \caption{[$\alpha$/Fe]–[Fe/H] diagram for the adopted subgiant sample, showing a significant bimodality, with the low-$\alpha$ thin-disk and high-$\alpha$ thick-disk sequences. High-$\alpha$ sequence stars were then selected using
$[\alpha/\mathrm{Fe}] > - 0.12\times[\mathrm{Fe/H}] + 0.1$,
$[\alpha/\mathrm{Fe}] > 0$, and
$[\mathrm{Fe/H}] > -1.5$ (red lines). Stars below the red lines are dominated by the traditional thin-disk population \citep{Imig_2023bimodal}.} 
    \label{fig:1}
\end{figure}

\begin{figure*}[t]
    \centering
        \includegraphics[width=0.73\textwidth]{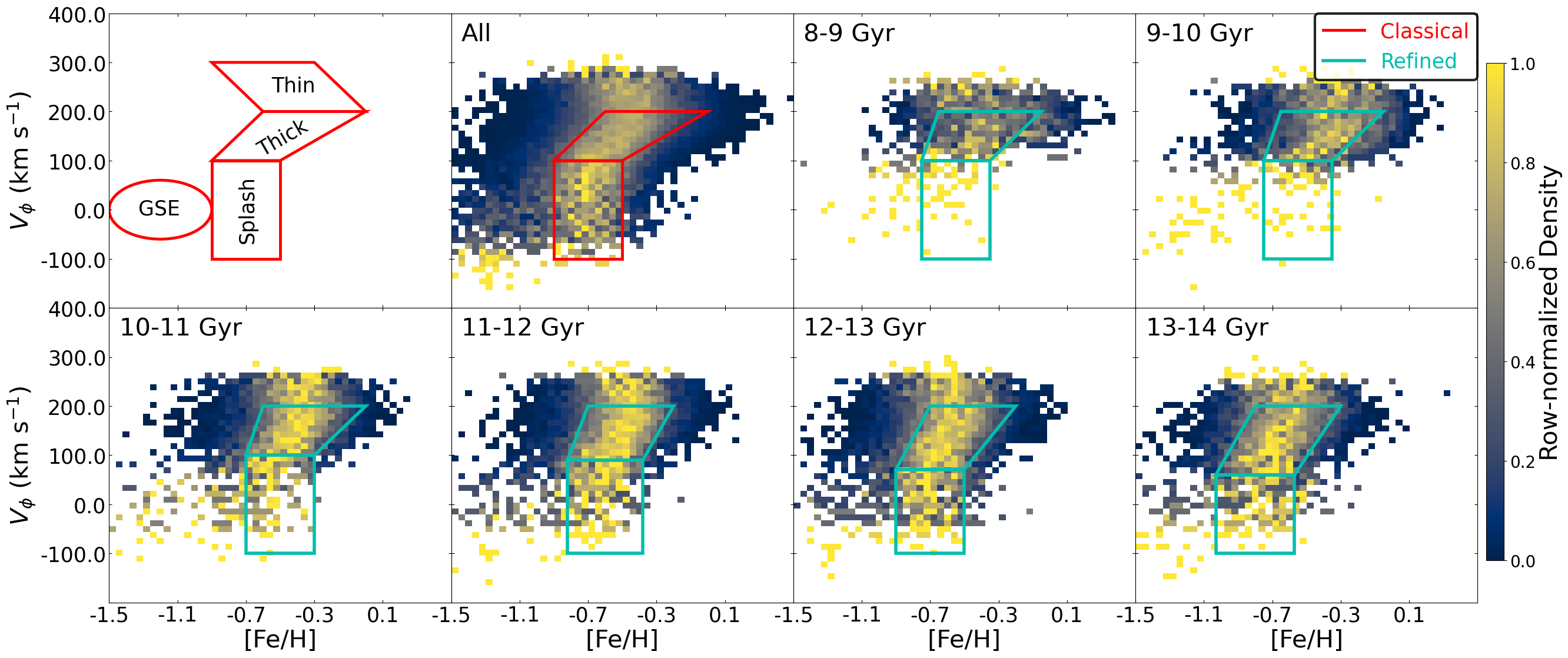}
        \includegraphics[width=0.22\textwidth]{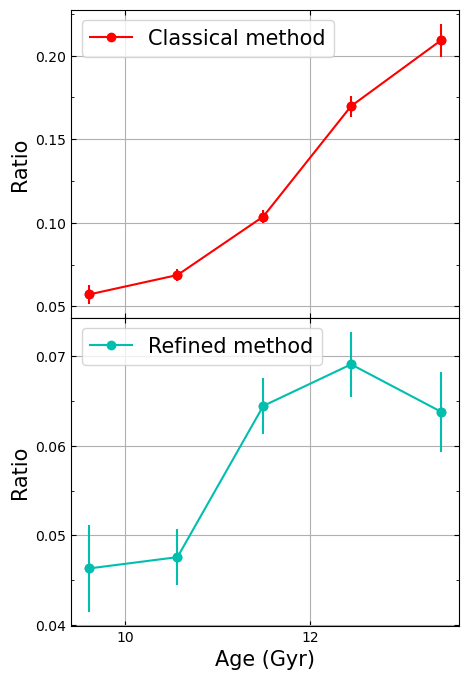}
    \caption{\textbf{Left:} Row-normalized rotation-velocity--[Fe/H] diagrams comparing the classical selection (red) with the refined criteria (green). The first panel shows the population boundaries from \citet{2020MNRASselect_splash}, and the second panel presents the distribution of all high-$\alpha$ subsample. The remaining panels show the distributions in different stellar age bins. In the oldest bin (bottom-right), the refined criteria adopt a lower metallicity threshold than the classical method, and the upper velocity limit of the Splash selection falls below 100 km s$^{-1}$. Toward younger ages, the refined Splash selection shifts to higher metallicity and higher rotational velocity.
\textbf{Right:} Number ratios of Splash to thick-disk stars as a function of stellar age. With the classical selection (red), the ratio declines steadily beyond 9 Gyr. In contrast, the refined method (green) yields an approximately constant ratio at the oldest ages, consistent with our assumption that Splash stars were produced with similar probability across stellar ages.}
     \label{fig:2}
\end{figure*}

\subsection{Sample selection} \label{sec:2.2}
\par We first applied chemical selection cuts to isolate stars belonging to the Milky Way thick disk, including the Splash population. The [$\alpha$/Fe]–[Fe/H] distribution (Figure~\ref{fig:1}) exhibits a clear bimodality , where $\alpha$ abundance refers to Mg, Si, Ca, and Ti in \citetalias{2022Nature_Xiang}. Following \citet{2019ApJS_xiang}, [$\alpha$/Fe] is defined as the inverse-variance-weighted mean of [Mg/Fe], [Si/Fe], [Ca/Fe], and [Ti/Fe], with weights $\omega_X=1/\sigma^2_{[X/{\rm Fe}]}$. Unless otherwise stated, [$\alpha$/Fe] in this work refers to this definition. We select the high-$\alpha$ sequence (stars above the red lines as defined in Figure~\ref{fig:1}), which predominantly traces the thick disk at intermediate metallicities, while the low-$\alpha$ sequence corresponds to the thin disk. In defining the high-$\alpha$ parent sample, we revise the classical [$\alpha$/Fe]--[Fe/H] boundary (e.g. \citealt{Imig_2023bimodal}) to minimize contamination from both low-$\alpha$ thin-disk stars and accreted halo stars before applying the kinematic selection. Specifically, the adopted boundary avoids the chemically intermediate region between the low- and high-$\alpha$ sequences, thereby reducing potential contamination from low-$\alpha$ disk stars scattered by measurement uncertainties, and imposes a stricter cut at the metal-poor end, where accreted halo stars are more likely to overlap with the high-$\alpha$ disk. After the subsequent $V_\phi$--[Fe/H] selection, the final thick-disk and Splash samples are restricted to [Fe/H] $>-1.1$. Using APOGEE DR17, we verify that this boundary does not introduce significant accreted-halo contamination in the [Mg/Mn]--[Al/Fe] plane \citep{2015_Hawkins,2020_Das}. We then separate the Splash and thick-disk stars in the $V_\phi$--[Fe/H] plane, as described below.\par 

Within the high-$\alpha$ sample, we further separated thick-disk and Splash populations using kinematic and metallicity cuts \citep{2020MNRASselect_splash}. The row-normalized $V_{\phi}$–[Fe/H] diagram (Figure~\ref{fig:2}, upper left) reveals three components:
(i) the thin disk, with $V_{\phi} > 200$ km s$^{-1}$ and a negative slope in the $V_{\phi}$–[Fe/H] relation consistent with epicyclic motions and the radial metallicity gradient \citep{Lee_2011};
(ii) the thick disk, with $100 < V_{\phi} < 200$ km s$^{-1}$ and a positive slope reflecting its gradual enrichment and spin-up \citep{Selecting_accreted_populations};
(iii) the Splash, characterized by lower (or even retrograde) rotation but metallicities comparable to those of the thick disk.
The presence of the thin-disk component is not expected in a purely high-$\alpha$ sample and likely reflects a small fraction of thin-disk stars scattered into the high-$\alpha$ region due to observational uncertainties.

\par The row-normalized $V_{\phi}$–[Fe/H] distributions in different age bins are shown in the remaining panels of Figure~\ref{fig:2}. Applying the same fixed boundaries at all ages introduces systematic biases. The thick disk evolves with time, becoming more metal-rich and rotating faster as age decreases \citep{2016_thick_metallicity_rich,2022_thick_rotation}. Consequently, fixed selection cuts may misclassify some thick-disk stars with low $V_{\phi}$ as Splash, while missing others due to shifts in the metallicity distribution.

To account for this evolution, we introduced age-dependent selection boundaries (green lines in Figure~\ref{fig:2}). Compared with the classical method, which assumes a time-independent thick-disk distribution, this refined selection incorporates the observed age dependence of the disk’s chemical and kinematic properties. Following \citet{2020MNRASselect_splash}, stars within the rectangular region are classified as Splash, while those above the Splash sequence (with a positive slope at the same metallicity) are assigned to the thick disk, keeping the upper velocity cutoff fixed at 200 km s$^{-1}$.

We consider the refined definition to better reflect the physical evolution of the thick disk and therefore adopt it as the fiducial sample for the main analysis. For comparison, results obtained using the classical selection are presented in Appendix~\ref{appendix:B}. The final samples contain 27,726 thick-disk stars and 1,841 Splash stars under the refined criteria, compared to 26,496 thick-disk stars and 3,446 Splash stars using the classical selection.

\section{Results} \label{sec:3}
In this section, we examine the age distributions of the Splash and thick-disk populations selected according to the refined definition described above. We first use their relative age-dependent trends to obtain an empirical estimate of the epoch of the GSE merger, as the changing fraction of Splash and thick-disk stars with age traces the transition induced by the merger. We then explain the apparent systematic age offset, whereby Splash stars appear older than thick-disk stars despite being interpreted as originating from the pre-merger thick disk. Finally, we constrain the merger epoch by modeling the age difference between the two populations, explicitly accounting for age uncertainties and adopting a Gaussian form for the star formation history (SFH) of the thick-disk population.

\subsection{Number Ratio} \label{sec:4.1} 
Before the GSE merger, the stars that now constitute the Splash population were part of the Galactic thick disk. Given that the thick disk formed over a relatively short timescale, its stars exhibit only modest variations in angular momentum and other intrinsic properties across different ages. We therefore assume that, at the time of the merger, disk stars of different ages were approximately equally likely to be dynamically heated into the Splash. Under this assumption, when traced back in time, the ratio of Splash to thick-disk stars prior to the merger should remain approximately constant with stellar age.\par

The right panels of Figure~\ref{fig:2} compare this ratio under the classical and refined selection schemes. Using the classical criteria, the ratio declines steadily from 14 to 9 Gyr, contrary to the expectation of a constant value. This behavior arises because, at the earliest epochs, the boundary separating thick-disk and Splash stars should lie at lower rotational velocity than in the well-established thick disk. As a result, the classical criteria tend to misclassify a fraction of thick-disk stars as Splash stars. As the thick disk spins up and becomes more metal-rich over time, this misclassification gradually decreases, producing the apparent decline in the ratio. In contrast, the refined criteria yield an approximately constant ratio between 11 and 14 Gyr, in better agreement with the assumption of age-independent Splash production.\par

Importantly, this conclusion is based on a completely independent diagnostic, the constancy of the Splash-to-disk number ratio, thereby independently confirming previous estimates of the merger epoch and supporting the robustness of our refined selection method.\par

\begin{figure}[t]
    \centering
    \plotone{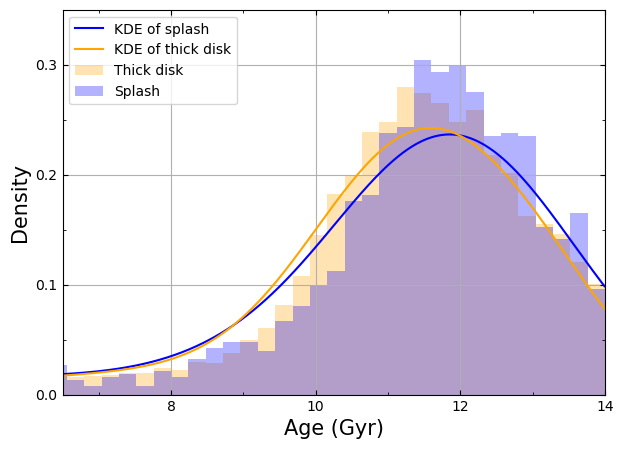}
    \caption{Age distribution histograms and their corresponding KDEs for the thick disk (orange) and Splash (blue) populations. The majority of stars in both components have ages older than 10 Gyr, indicating that the present-day thick disk formed at early times. Nevertheless, the Splash population shows a relatively higher fraction at the oldest ages compared to the thick disk.}
    \label{fig:3}
\end{figure}

\begin{figure}
    \centering
    \plotone{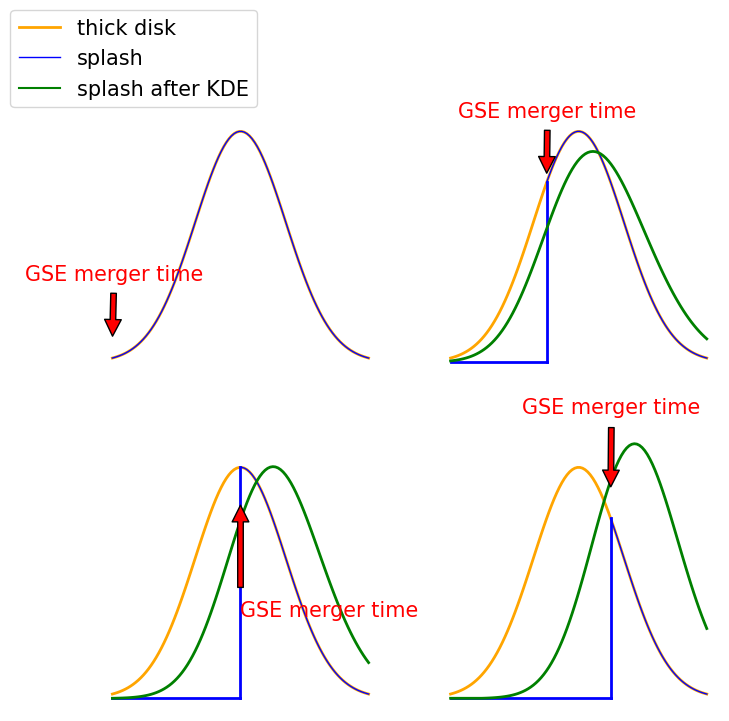}
    \caption{Illustration of why the Splash appears older than the thick disk. We assume that the thick-disk star-formation history (orange) follows a Gaussian distribution. Prior to the GSE merger, the intrinsic age distribution of Splash stars (blue) is identical to that of the thick disk, but Splash formation is truncated at the merger epoch. The Splash curve therefore drops abruptly at the merger time (blue step). Convolution with Gaussian age uncertainties produces the “observed” Splash distribution (green), in which the peak shifts toward older ages due to the lack of the younger part of the distribution. The four panels show different assumed merger epochs, from top left to bottom right: near the end of thick-disk formation, just after the peak, around the peak, and before the peak. The combination of truncation and age uncertainties makes the Splash appear older than the thick disk.}
    \label{fig:4}
\end{figure}

\subsection{Peak shift of age distribution} \label{sec:4.2}

The age distributions of the thick-disk and Splash stars, together with their kernel density estimates (KDEs), are shown in Figure~\ref{fig:3}. Both populations are very old; however, in both the distributions and their KDEs, the Splash appears systematically older than the thick disk. This is puzzling: if the Splash indeed originated from the thick disk and was dispersed by the merger, one would not expect it to predate its progenitor. Moreover, the peak of the Splash distribution is clearly offset from that of the thick disk, further highlighting the discrepancy.\par

To investigate this apparent offset, we illustrate the effect of age uncertainties in Figure \ref{fig:4}. Before the merger, the intrinsic age distributions of the Splash (blue) and thick disk (orange) should have been identical. After the merger, Splash formation was truncated, producing a distribution with a sharp cutoff at the merger time, while the thick disk did not undergo this change and continued to form stars. Convolution with Gaussian age errors shifts the apparent Splash peak to older ages (green): while a full Gaussian distribution retains its peak position under convolution, a truncated one loses density on the younger side, displacing the maximum to the right, which is precisely the Eddington-like bias, a skewed distribution caused by measurement errors. This naturally explains why the Splash distribution appears older than the thick disk, even if both originated from the same underlying population.\par

\subsection{GSE merger time} \label{sec:4.3}
In the above section, we demonstrated that uncertainties in stellar age estimates can induce a peak shift in the Splash age distribution. Comparing the different cases in Figure~\ref{fig:4}, we find that the impact of the GSE merger on the peak shift depends on its timing relative to the peak of thick-disk star formation. In particular, mergers occurring earlier, i.e., prior to the peak of the thick-disk SFH, produce a larger separation between the Splash and thick-disk age peaks. Quantitatively,
\begin{equation}
\Delta t = t_{\mathrm{splash}}^{\mathrm{peak}} - t_{\mathrm{thick}}^{\mathrm{peak}} \sim f\!\left(t_{\mathrm{merger}}, t_{\mathrm{thick}}^{\mathrm{peak}}\right).
\end{equation}
Because these two quantities are monotonically related, the observed value of $\Delta t$ can be used to constrain the epoch of the GSE merger.\par

From the KDE in Figure~\ref{fig:3}, we measure the peak of thick-disk star formation at $11.58$~Gyr. We note that the original catalog from \citetalias{2022Nature_Xiang} adopts an upper age limit of 20~Gyr; within this range, the age distribution is well described by a Gaussian with a peak at $11.59$~Gyr. This value is nearly identical to our result after excluding stars older than the age of the Universe (13.8~Gyr), indicating that the peak measurement is robust against this cut. Therefore, although the age distribution in Figure~\ref{fig:3} deviates slightly from a perfect Gaussian, its peak can still be reliably determined.
We estimate $\Delta t$ and its uncertainty via Monte Carlo sampling. Specifically, we draw 10{,}000 realizations of stellar ages accounting for individual errors, determine the peak of each realization, and construct distributions of peak ages for both the Splash and thick-disk populations. From these, we obtain
\[
\Delta t = 0.30^{+0.06}_{-0.06}\,\mathrm{Gyr}, \quad 
t_{\mathrm{thick}}^{\mathrm{peak}} = 11.58^{+0.02}_{-0.02}\,\mathrm{Gyr}.
\]\par

\begin{figure}[t]
    \centering
        \plotone{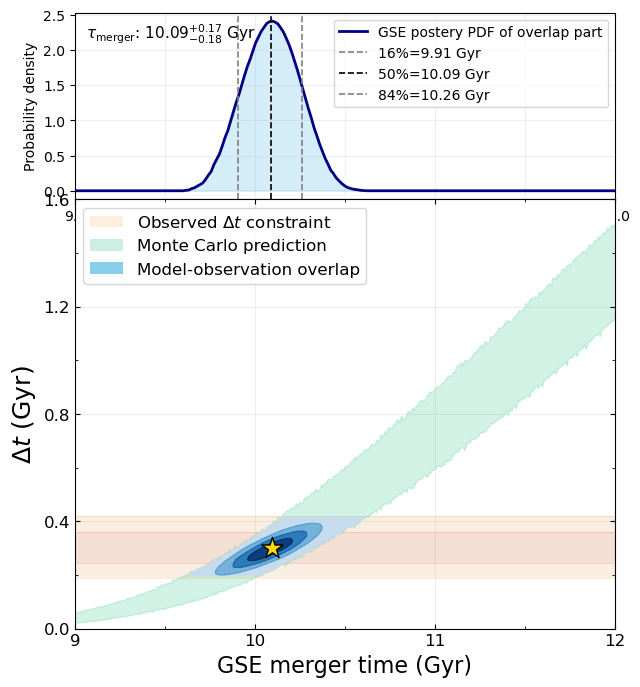}
    \caption{ 
   Joint constraint on the GSE merger epoch from the observed peak offset. The green band shows the $2\sigma$ $\Delta t$–$t_{\mathrm{GSE}}$ relation from Monte Carlo simulations, while the orange band indicates the $2\sigma$ observational constraint on $\Delta t$. Their intersection defines the region consistent with the data (blue contours), with the star marking the maximum-likelihood solution. The upper panel shows the posterior distribution of the merger epoch.
    }
    \label{fig:5}
\end{figure}

Adopting the observed peak of the thick-disk SFH as a prior, we model the effect of different GSE merger times on the expected peak offset. We first construct an intrinsic age distribution for the thick disk by assuming a finite width, estimated by subtracting the median age uncertainty in quadrature from the observed distribution. 
For a given merger epoch, we generate the corresponding intrinsic Splash age distribution by assuming that stars are dynamically heated into the Splash at the time of the merger. This intrinsic distribution is then convolved with the age uncertainties to obtain the observed Splash age distribution. From this, we measure the expected peak offset, defined as the difference between the peak of the thick-disk SFH and that of the observed Splash age distribution, $\Delta t$.
Repeating this procedure over a range of merger times, we derive the functional relation between the peak offset and the merger epoch, $\Delta t \sim f(t_{\mathrm{merger}} \mid t_{\mathrm{thick}}^{\mathrm{peak}})$, as shown by the green region in Figure~\ref{fig:5}. Combining this relation with the observational constraint on $\Delta t$, we obtain the posterior distribution of the GSE merger time, shown in the upper panel of Figure~\ref{fig:5}. The merger epoch is inferred to be $t_{\mathrm{GSE}} = 10.1^{+0.2}_{-0.2}$~Gyr ago.

This estimate is consistent with the number-ratio analysis presented in Section~\ref{sec:4.1} and with previous determinations based on stellar kinematics and chemical abundances \citep[e.g.,][]{Helmi2018,2018_haywood,2021_Naidu}. However, the constraint derived here is significantly tighter, reducing the allowed time interval of the GSE merger to $\sim0.4$~Gyr. Such a narrow window provides a precise anchor for the evolutionary timeline of the Milky Way.

\section{Discussion and Conclusion} \label{sec:4}

To constrain the epoch of the GSE merger, we adopt the simplifying assumption that the merger dynamically perturbed thick-disk stars with equal probability across all ages, thereby producing the Splash population whose age distribution, at the moment of the merger, mirrors that of the contemporaneous thick disk \citep{2020MNRASselect_splash,2025_affect_motion}. The thick-disk SFH is modeled as a Gaussian distribution, and the merger is treated as an instantaneous event that truncates the Splash age distribution. 
We employ two complementary approaches to constrain the merger epoch. Both are subject to uncertainties in the adopted criteria for separating Splash and thick-disk stars, but each offers distinct advantages.

The number-ratio method is largely insensitive to observational incompleteness, since Splash and thick-disk stars are drawn from the same parent catalog and selection effects largely cancel in the ratio. Although this method does not explicitly account for age uncertainties, we show in Appendix~\ref{appendix:A} that their smoothing effect cannot reproduce the behavior seen in the classical selection (right panel of Figure~\ref{fig:2}); even after accounting for age uncertainties, the ratio remains approximately constant with age at the old end.

The peak-offset method, in contrast, yields a substantially tighter constraint on the merger epoch, improving the precision by a factor of $\sim5$--10 compared to previous estimates (e.g., \citealt{2021_Naidu}). We also verify that the apparent offset between the peaks of the Splash and thick-disk age distributions persists under the classical selection, confirming that this offset is not introduced by the refined selection method. The refined criteria primarily improve Splash sample purity by reducing thick-disk contamination, a conclusion independently supported by the number-ratio analysis. In practice, the stricter rotation-velocity cut removes part of the oldest tail of the Splash distribution, making the refined sample slightly younger than the classical one. This reduces the peak offset relative to the thick-disk SFH peak and shifts the inferred merger epoch to a later time. Results based on the classical selection are presented in Appendix~\ref{appendix:B}.
Here we note that the Gaussian SFH assumed for the thick disk is adopted for simplicity; a more realistic description may further refine the inferred merger epoch, which we leave for future work. We also emphasize that a proper estimate of age uncertainties is essential for robustly inferring the merger epoch. Recent work \citep{2025arXiv251008675S}, based on wide binaries, suggests that the age uncertainties of subgiants in \citetalias{2022Nature_Xiang} are well characterized.

The GSE sample itself provides an additional, independent perspective. If star formation in the GSE progenitor was also strongly suppressed by the merger, its present-day age distribution should encode the same event. Using a dynamical selection in the $J_r$--$L_z$ plane \citep{Selecting_accreted_populations,2022_Buder,2023_Gaia_col,2024_GSE_selection}, we find that the GSE age distribution peaks at 11.92~Gyr, compared with 11.86~Gyr for the Splash population. Both peaks are older than that of the thick disk (Appendix~\ref{appendix:C}). Notably, the GSE peak is also older than that of the Splash population. Since both populations are subject to similar Eddington-like biases, this suggests that the intrinsic peak of the GSE SFH is likely older than that of the Galactic thick disk.
However, a quantitative inversion remains challenging, as the intrinsic width and detailed form of the GSE age distribution are uncertain.

We note that Splash-like populations may also arise through alternative channels, including scattering by massive clumps in an early clumpy disk, proto-disk stars born on dynamically hot orbits, and secular processes such as bar resonances \citep{2020_Amarante,2023_Dillamore,2026_buder}. Some of these channels may also help explain the observed age offset between the Splash and the high-$\alpha$ disk; for example, clump scattering could preferentially place older, dynamically hotter stars onto low-angular-momentum Splash-like orbits, producing an older age distribution than that of the later high-$\alpha$ disk \citep{2020_Amarante,2025_Amarante}. Whether these mechanisms operated in the Milky Way remains uncertain and requires further investigation, which is beyond the scope of this Letter. We therefore conduct our analysis within the framework of the standard scenario in which the Splash was formed through GSE heating.

In summary, we first introduce a refined Splash selection method that yields a cleaner Splash sample, as independently supported by the number-ratio analysis. The latter also provides an independent estimate of the merger epoch at $\sim10$--11~Gyr. We then quantify the apparent age offset between the Splash and thick-disk populations and show that it arises primarily from observational biases, specifically the combination of truncation imposed by the merger and measurement uncertainties, rather than from a genuinely older Splash population. Using this effect, we constrain the GSE merger epoch to a narrow window of $10.1^{+0.2}_{-0.2}$~Gyr. 
Although our analysis adopts simplified assumptions regarding thick-disk evolution and the nature of the merger, the combination of complementary diagnostics provides new and independent insight into the origin of the Splash substructure and the imprint of the GSE event on the Galactic disk. Looking ahead, forthcoming datasets such as Gaia DR4 and large-scale asteroseismic surveys will deliver more precise stellar ages for vast samples, enabling tighter constraints on the timing of the GSE merger and more stringent tests of our framework.

\section*{Acknowledgments}
We thank the anonymous referee for constructive comments. This work acknowledges the supports from National Key R\&D Programme of China (Grant No. 2025YFF0510603, 2024YFA1611903, and 2023YFA1608303) and the National Science Foundation of China (NSFC Grant No. 12422303) and the Fundamental Research Funds for the Central Universities (Grant No. 118900M122, E5EQ3301X2 and E4EQ3301X2). This work has made use of data from the European Space Agency (ESA) mission {\it Gaia} (\url{https://www.cosmos.esa.int/gaia}), processed by the {\it Gaia} Data Processing and Analysis Consortium (DPAC, \url{https:// www.cosmos.esa.int/web/gaia/dpac/ consortium}). Funding for the DPAC has been provided by national institutions, in particular the institutions participating in the {\it Gaia} Multilateral Agreement.
This work made use of the data from LAMOST (Large Sky Area Multi-Object Fiber Spectroscopic Telescope, also known as the Guoshoujing Telescope) (https://cstr.cn/31118.02.LAMOST). LAMOST is a Chinese national mega-science facility, operated by National Astronomical Observatories, Chinese Academy of Sciences.

\bibliographystyle{aasjournal}
\bibliography{sample631}{}


\appendix
\setcounter{table}{0}   
\setcounter{figure}{0}
\renewcommand{\thetable}{A\arabic{table}}
\renewcommand{\thefigure}{A\arabic{figure}}

\section{Impact of error on number ratio}\label{appendix:A}
\begin{figure}[htbp!]
    \centering
    \includegraphics[width=0.4\textwidth]{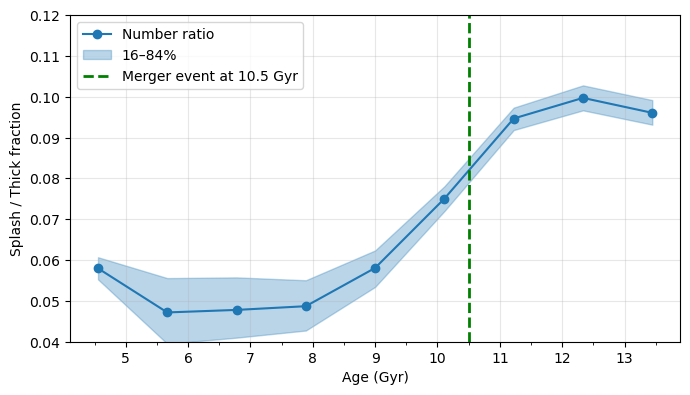}
    \caption{
    Age dependence of the Splash-to-thick-disk number ratio under our merger-ejection toy model. 
    The green dashed line marks the adopted merger time at 10.5~Gyr. 
    Assuming that 10\% of thick-disk stars are transferred into the Splash population during the merger, we estimate the Splash/thick-disk fraction in each age bin by Monte Carlo resampling the stellar ages 1000 times according to their age uncertainties. 
    The blue solid line shows the median ratio, and the blue shaded region indicates the 16th--84th percentile range.
}
    \label{fig:6}
\end{figure}
To test whether age uncertainties could erase the expected pre-merger signature, we construct a simple merger-ejection model in which the Splash population is drawn from the thick disk at 10.5~Gyr with an age-independent ejection probability of 10\%. 
We additionally include a 5\% post-merger contamination of thick-disk stars into the Splash sample to mimic observational errors and classification effects, while requiring at least 10 Splash stars in each age bin, consistent with the observed Splash age distribution in Figure~\ref{fig:3}. 
We then resample the age of each thick-disk and Splash star 1000 times according to its age uncertainty and recompute the Splash/thick-disk number ratio. 
As shown in Figure~\ref{fig:6}, the ratio remains nearly flat before the merger, indicating that the relatively small age uncertainties do not wash out the pre-merger plateau; this contrasts with the smoother trend obtained with the classical method in Figure~\ref{fig:3}.

\section{Results using Splash sample selected from classical method}\label{appendix:B}

\begin{figure}[htbp!]
    \centering
    \includegraphics[width=0.4\textwidth]{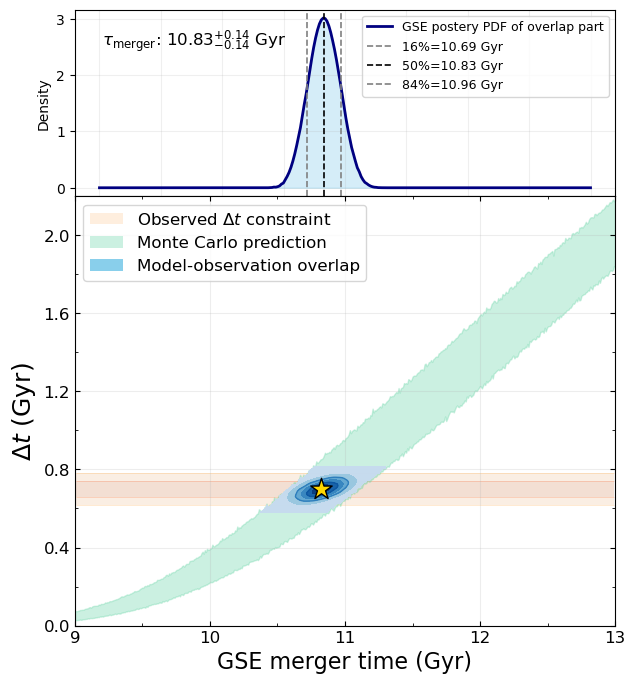}
    \caption{
    Constraint on the epoch of the GSE merger derived from the classical Splash and thick-disk selection. 
    The green band shows the $2\sigma$ $\Delta t$--$t_{\mathrm{GSE}}$ relation predicted by the Monte Carlo simulations, and the orange band indicates the $2\sigma$ observational constraint on the peak offset, $\Delta t = 0.70^{+0.04}_{-0.04}$~Gyr. 
    Blue contours represent the model--observation overlap density, and the star marks the maximum-likelihood solution. 
    The upper panel shows the marginalized probability distribution of the inferred merger epoch, yielding $t_{\mathrm{merger}} = 10.83^{+0.14}_{-0.14}$~Gyr.
}
    \label{fig:7}
\end{figure}

To assess the sensitivity of the inferred merger epoch to the sample definition, we repeat the analysis using the classical Splash and thick-disk selection (Figure~\ref{fig:7}).  The preferred value shifts to $t_{\mathrm{merger}} = 10.83^{+0.14}_{-0.14}$~Gyr, but remains qualitatively consistent with the result obtained in the main text, indicating that the $\Delta t$-based inference remains applicable under the classical selection.

The difference is also understandable. Compared with the refined selection adopted in the main text, the classical method admits a wider range of rotational velocities, especially at the old-age end. It therefore suffers more seriously from contamination by older stellars. As a result, the Splash age distribution is shifted toward older ages, and the peak offset, $\Delta t$, with respect to the thick disk becomes larger. In the framework of our model, the merger epoch is related to $\Delta t$. A larger $\Delta t$ therefore leads to an older inferred merger time, consistent with the trend discussed in Section~\ref{sec:4.2}.

This result therefore provides a useful consistency check: even under the classical selection, the peak-offset method still recovers the same basic picture that the Splash is older than the thick disk. However, the classical selection is more vulnerable to contamination from older populations, which increases the observed peak offset and biases the inferred merger epoch toward older ages. Moreover, the number-ratio analysis independently favors the refined selection. We therefore consider the refined selection in the main text to be both physically better motivated and more reliable for estimating the merger time.

\section{Age distribution of GSE stars}\label{appendix:C}
To construct a comparison sample of GSE stars, we adopt a dynamical selection rather than a purely chemical one. 
The reason is that our catalog includes only Mg and Fe abundances, which do not allow us to apply more restrictive chemical-selection criteria such as those based on [Mg/Mn] and [Al/Fe]. 
We therefore select GSE stars in the $\sqrt{J_r}$--$L_z$ plane.

\begin{figure*}[htbp!]
    \centering
    \includegraphics[width=0.45\textwidth]{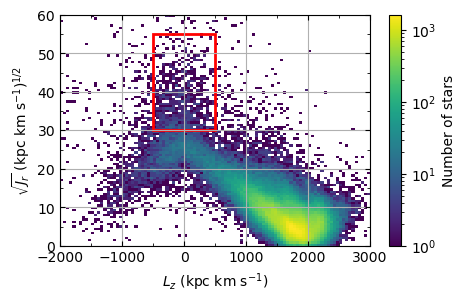}
    \hfill
    \includegraphics[width=0.45\textwidth]{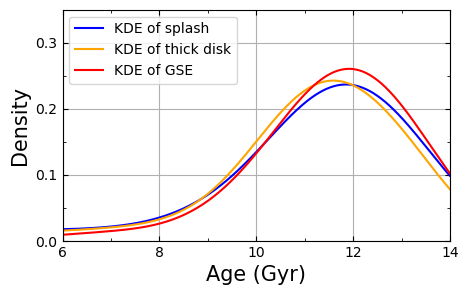}
    \caption{
    Left: Dynamical selection of the GSE sample in the $\sqrt{J_r}$--$L_z$ plane. 
    The red box marks the adopted selection region, defined by $30 \leq \sqrt{J_r} \leq 55$~(kpc km s$^{-1}$)$^{1/2}$ and $-500 \leq L_z \leq 500$~kpc km s$^{-1}$. 
    Right: KDE-smoothed age distributions of the thick-disk, Splash, and GSE samples, with stellar age uncertainties taken into account. 
    The peak ages of the thick-disk, Splash, and GSE populations are 11.59, 11.86, and 11.92~Gyr, respectively.}
    \label{fig:8}
\end{figure*}

Among dynamical selection methods, the $\sqrt{J_r}$--$L_z$ approach is widely used to obtain a relatively clean GSE sample. 
Accordingly, we define the GSE population using the selection box shown in the left panel of Figure~\ref{fig:8}. 

The right panel shows the KDE-smoothed age distributions of the thick-disk, Splash, and GSE samples after propagating the age uncertainties. 
Their peak ages are 11.59, 11.86, and 11.92~Gyr, respectively, indicating that the Splash and GSE populations have very similar characteristic ages.

\end{document}